# Anomalous Hall effect in two-dimensional non-collinear antiferromagnetic semiconductor Cr$_{0.68}$Se

J. Yan[1,2], X. Luo[1*], F. C. Chen[1,2], Q. L. Pei[1], G. T. Lin[1,2], Y. Y. Han[3], L. Hu[1], P. Tong[1], W. H. Song[1], X. B. Zhu[1] and Y. P. Sun[3,1,4*]

[1] *Key Laboratory of Materials Physics, Institute of Solid State Physics, Chinese Academy of Sciences, Hefei, 230031, China*

[2] *University of Science and Technology of China, Hefei, 230026, China*

[3] *High Magnetic Field Laboratory, Chinese Academy of Sciences, Hefei, 230031, China*

[4] *Collaborative Innovation Center of Advanced Microstructures, Nanjing University, Nanjing, 210093, China*

**Abstract**

Cr$_{0.68}$Se single crystals with two-dimensional (2D) character have been grown, and the detailed magnetization $M(T)$, electrical transport properties (including longitudinal resistivity $\rho_{xx}$ and Hall resistivity $\rho_{xy}$ and thermal transport ones (including heat capacity $C_p(T)$ and thermoelectric power (TEP) $S(T)$) have been measured. There are some interesting phenomena: (i) Cr$_{0.68}$Se presents a non-collinear antiferromagnetic (AFM) semiconducting behavior with the Néel temperature $T_N$ = 42 K and the activated energy $E_g$=3.9 meV; (ii) It exhibits the anomalous Hall effect (AHE) below $T_N$ and large negative magnetoresistance (MR) about 83.7% (2 K, 8.5 T). The AHE coefficient $R_S$ is 0.385 cm$^{-3}$/C at $T$=2 K and the AHE conductivity $\sigma_H$ is about 1 Ω$^{-1}$cm$^{-1}$ at $T$=40 K, respectively; (iii) The scaling behavior between the anomalous Hall resistivity $\rho_{xy}^A$ and the longitudinal resistivity $\rho_{xx}$ is linear and further analysis implies that the origin of the AHE in Cr$_{0.68}$Se is dominated by the skew-scattering mechanism. Our results may be helpful for exploring the potential application of these kind of 2D AFM semiconductors.

*Corresponding authors: xluo@issp.ac.cn and ypsun@issp.ac.cn

Since the Anomalous Hall effect (AHE) is found in ferromagnetic (FM) conductors [1], it has been a long-standing topic in condensed matter physics. AHE has been attracting significant interest due to their highly tunable physical properties and immense potential in spintronic device applications [2-6]. AHE is that the Hall resistivity is proportional to the magnetization and it is usually observed in ferromagnets [7-10]. However, since intrinsic AHE arises owing to Berry curvature induced by the external field [11], it may also appear in antiferromagnets [12-15]. Compared with the ordinary Hall effect (OHE), originating from the deflection of charge carriers by the Lorentz force in a magnetic field, the origin of the AHE is more complex. So far, three mechanisms responsible for AHE have been widely accepted. The intrinsic Karplus and Luttinger (KL) mechanism is related to the spin-orbit coupling (SOC) and perturbation by applying an electric field, resulting in an additional term to the usual group velocity [7, 16]. Induced by asymmetric scattering of the conduction electrons or impurities which are subject to SOC, the extrinsic mechanisms involving skew-scattering mechanism and side-jump one give rise to the AHE [17, 18]. Usually, the scaling relationship between the anomalous Hall resistivity $\rho_{xy}^A$ and the longitudinal resistivity $\rho_{xx}$ can be used to distinguish the different mechanisms.

Recently, transition-metal dichalcogenides (TMDs) have attracted significant attention. Due to the two dimensional (2D) character, the charge and heat transport are usually confined to the intralayer, and then, lots of unusual physical phenomena emerge in TMDs [19–25]. However, the AHE is rarely observed in TMDs due to seldom appearance of magnetism. The typical one, $Fe_{1/4}TaS_2$, presents the AHE below the FM temperature $T_C$=160 K with hedgehog spin texture [26]. For the TMDs with complex magnetic structure, to explore the origin of AHE, one also needs to consider the external interactions, such as, the interaction between the localized spins and the applied magnetic field. Therefore, the origin of the AHE in $Fe_{1/4}TaS_2$ still has not been understood well so far. In order to explore the origin of AHE in TMDs, it is valuable to explore more TMDs with AHE, which present different ground state nature and crystal structure. The chromium-selenium system $Cr_xSe$ (x<1) contain many binary compounds with different structural phases. For 1>x>0.87, it is the α phase with a hexagonal NiAs-type (B8) structure. The β phase with 0.85>x>0.71 crystallizes into a monoclinic phase. For 0.7>x>0.68, it is the γ phase with hexagonal symmetry. Although $Cr_xSe$ compounds have different structural phase, they can be described in terms of crosslinks of type $(1/y)Cr^{y+}$ between $[CrSe_2]_n^{n+}$ layers. For example, $Cr_2Se_3$ can be represented to $[CrSe_2]^-1/3Cr^{3+}$, and $[CrSe_2]^-1/2Cr^{2+}$ for $Cr_3Se_4$ and so on [27]. Although some $Cr_xSe$ compounds have been synthesized so far, such as $Cr_2Se_3$, $Ce_3Se_4$, $Cr_7Se_8$ and $Cr_{0.68}Se$, there are lots of researches on the magnetism of $Cr_2Se_3$ [28-30], $Cr_3Se_4$ [28], $Cr_7Se_8$ [31]. $Cr_2Se_3$ and $Cr_3Se_4$ exhibit antiferromagnetic (AFM) ground state, while $Cr_7Se_8$ shows spin-glass-like behavior. Although $Cr_{0.68}Se$ single crystals have been synthesized, the physical properties of $Cr_{0.68}Se$ have been rarely reported. For this purpose, we grew the $Cr_{0.68}Se$ single crystals and did the detailed research on the physical properties using magnetic, electronic and thermal transport measurements. The AHE is observed below the AFM temperature ($T_N$), and the linear scaling behavior between $\rho_{xy}^A$ and $\rho_{xx}$ is obtained and analysis results implies that the skew-scattering mechanism dominates the AHE in $Cr_{0.68}Se$. Our results may help to understand the origin of AHE in the other 2D AFM semiconductors.

The experimental details, the crystal structure and X-ray diffraction (XRD) of single and crushed crystal can be found in **supporting materials**.

In order to systematically investigate the basic physical properties of the $Cr_{0.68}Se$ single crystal, we carried out the measurements of magnetization as a function of temperature $M(T)$ and magnetic field $M(H)$ along $ab$ plane and $c$ axis. Figures 1 (a) and (b) show $M(T)$ under zero-field-cooling (ZFC) and field-cooling (FC) modes with the applied field $H$ = 0.05 T along $ab$ plane and $c$ axis, respectively. The Néel temperature $T_N$ of $Cr_{0.68}Se$ is around 42 K and the thermal magnetic irreversible phenomenon has been observed and the temperature $T_{irr}$ is around 15 K (details shown in the inset of Figs. 1 (a) and (b)). In the paramagnetic (PM) state, the temperature dependence of susceptibility ($\chi(T)$) follows the Curie-Weiss law and the fitting curves are shown by blue solid lines in Figs. 1 (a) and (b). The Curie-Weiss law can be described as follows:

$$\chi(T) = \frac{M}{H} = \frac{C}{T - \theta_p} + \chi_0 \qquad (1)$$

where $C$ is the Curie constant, $\theta_p$ is the Weiss temperature and $\chi_0$ is the Pauli PM constant. From the analysis by Eq. (1), we can obtain the negative Weiss temperature $\theta_p$ ($H // c$) = -82.1 K and $\theta_p$ ($H // ab$) = -90.6 K, respectively, which indicates dominating AFM interaction along both $c$ and $ab$ directions. The Curie constant $C$ ($H // c$) = $C$ ($H // ab$) =1.23 emu K/mol.

The effective magnetic moment $\mu_{eff}$ can be calculated by:

$$\mu_{eff}/\mu_B = \sqrt{3k_B C/N_A Z} \qquad (2)$$

where $k_B = 1.38 \times 10^{-16}$ erg/K, $N_A = 6.02 \times 10^{23}$ mol$^{-1}$ and $Z$ is the atom number per unit cell. From the Eq. (2), we can get the effective moment of Cr ions in Cr$_{0.68}$Se:

$$\mu_{eff} = 3.13 \, \mu_B \tag{2.a}$$

The inset of Figs. 1 (a) and (b) exhibit the field dependence of magnetization $M(H)$ at various temperature between 2 and 300 K for $H // c$ axis and $H // ab$ plane, respectively. The magnetization of Cr$_{0.68}$Se shows a nonsaturated character under 4.5 T, which may be induced by the non-collinear AFM interaction of CrSe$_2$ layers in Cr$_{0.68}$Se [32-34]. And the $M(H)$ also exhibits isotropic property, indicating that spin canting should be considered. In Cr$_{0.68}$Se, there are two kind of Cr atom sites with different magnetic ordering. From the $M(T)$ curves along $ab$ plane, a kink is observed at $T_N$ and PM behavior is shown between $T_N$ and $T_{irr}$, which implies that the magnetic moments of intercalated Cr ions may be in PM arrangement above the $T_{irr}$. Below the $T_{irr}$, they tend to FM arrangement with spin canting. Therefore, the irreversibility of the ZFC and FCC curves may result from the competition between the AFM interaction of the CrSe$_2$ layers and the FM interaction of the intercalated Cr ions, and eventually results in the appearance of the spin-glass-like state [35], more detailed discussion on the magnetic structure of Cr$_{0.68}$Se can be found in the following text.

With regard to the transport properties of Cr$_{0.68}$Se single crystal, Fig. 1 (c) shows the temperature dependence of the resistivity along the $ab$ plane. Cr$_{0.68}$Se shows a semiconducting behavior because of the negative $d\rho/dT$. In the inset of Fig. 1 (c), $ln(\rho)$ decreases almost linearly with increasing $T$ at high temperature region, the activation energy can be calculated by using Arrhenius equation:[36]

$$ln\rho = ln\rho_0 + \frac{E_a}{k_B T} \tag{3}$$

As shown in the inset of Fig. 1 (c), the activation energy $E_a$ is about 3.9 meV, which indicates the Cr$_{0.68}$Se is a narrow-gap semiconductor.

To study the thermal property, we performed the specific heat measurement of Cr$_{0.68}$Se. Figure 1 (e) shows the temperature dependence of the heat capacity $C_P(T)$ for Cr$_{0.68}$Se. The $C_P(T)$ exhibits an anomaly around $T_N = 42$ K, which is consistent with the magnetization data. Debye model is used in heat capacity fitting, the $C_{mag}(T)$ can be calculated by the following equations: [36]

$$C_{V\,Debye}(T) = 9R\left(\frac{T}{\Theta_D}\right)^3 \int_0^{\Theta_D/T} \frac{x^4 e^x}{(e^x-1)^2} dx \quad (4a)$$

$$C_{mag}(T) = C_P(T) - nC_{V\,Debye}(T) \quad (4b)$$

while $n$ = 1.68 is the number of atoms per formula unit, $R$ is the molar gas constant and $\Theta_D$ is the Debye temperature. As shown in Fig. 1 (d), the red solid line is the calculated heat capacity using Debye model by Eq. (4a). The fitted Debye temperature $\Theta_D$ is about 312 K. The $C_{mag}(T)$ can be obtained by Eq. (4b) and the heat-capacity jumps at 42 K, which is shown in Fig. 1 (e). The magnetic entropy $S_{mag}$ is obtained by integrating the $C_{mag}/T$ versus $T$:

$$S_{mag}(T) = \int_0^T \frac{C_{mag}(T)}{T} dT \quad (5)$$

The temperature dependence of $S_{mag}$ is shown in Fig. 2 (f). The theoretical expected value of the magnetic entropy $S_{mag}$ is 4.9 J/mol K. It can been seen that the value of $S_{mag}$ is about 80% of the theoretical expected one. **Table 2S (supporting materials)** summarized all the parameters from the analysis above.

We further investigated the transport properties of $Cr_{0.68}Se$. Figure 2 (a) shows the magnetic field dependence of longitudinal resistivity $\rho_{xx}(H)$ at different fixed temperatures. At high temperature, the value of $\rho_{xx}(H)$ is nearly constant with increasing magnetic field. With decreasing temperature, the $\rho_{xx}(H)$ decreases dramatically under the high magnetic field, which indicates that large negative magnetoresistance (MR) is observed in $Cr_{0.68}Se$. Similar phenomena has been reported in anionic doping TMDs TiTeI, the intrinsically frustrated antiferromagnetism acted as a spin-dependent scattering barrier produced the large negative MR [32]. Moreover, the recent first-principle calculations predicted that $CrSe_2$ has the non-collinear AFM magnetic structure [37]. In our case, $Cr_{0.68}Se$ is isostructural with $CrSe_2$ except for few Cr ions intercalated into the $CrSe_2$ layers, the non-collinear AFM ordering of $CrSe_2$ layers may be suggested to the origin of the large negative MR in $Cr_{0.68}Se$. However, further studies are needed to prove this point.

Figure 2 (b) presents the Hall resistivity $\rho_{xy}(H)$ as a function of applied magnetic field $H$ for the $Cr_{0.68}Se$ at different temperatures. The $\rho_{xy}(H)$ increases quickly to certain saturated values at low $H$ region below $T_N$ = 42 K. With increasing $H$ further, the $\rho_{xy}(H)$ decreases slightly and the $H$ dependence of $\rho_{xy}(H)$ is almost linear, *i.e.*, the $\rho_{xy}(H)/H$ is constant, which clearly indicates that there is an AHE in $Cr_{0.68}Se$ [7, 8].

Conventionally, the Hall resistivity is described by: [8, 12-14]

$$\rho_{xy} = \rho_{xy}^O + \rho_{xy}^A = R_0 B + R_S \mu_0 M \qquad (6)$$

where $\rho_{xy}^O$, $\rho_{xy}^A$, $R_0$ and $R_S$ are the ordinary, AHE resistivity, the ordinary Hall coefficient and anomalous Hall coefficient, respectively. As shown in Fig. 2 (c) and (d), the values of $R_0$ and $\rho_{xy}^A$ in principle can be determined from linear fitting of the $\rho_{xy}(H)$ curves at saturation region. The slope and y-axis intercept is corresponding to the $R_0$ and $\rho_{xy}^A$, respectively. It shows that the dominant carrier in $Cr_{0.68}Se$ is electron because the values of $R_0$ are negative below the $T_N$, which is well agreed with the thermoelectric power (TEP) measurements, as shown in **Fig. 3S (supporting materials)**. From $R_0(T) = 1/nq$, the calculated carrier concentration $n$ is shown in inset of Fig. 2 (c). The $n_a(T)$ shows a moderate temperature dependence, the value of $R_0$ increases with the temperature below $T_N$, which may be induced by the spin reorientation of the $CrSe_2$ layers in $Cr_{0.68}Se$, leading to the change of carrier concentration [8].

The $R_S$ can be obtained by using the formula $\rho_{xy}^A = R_S \mu_0 M_S$, with the $M_S$ taken from $M(H)$ curves, which is shown in Fig. 2 (d). It shows that the values of $R_S$ decreases monotonically to minimum at 15 K, which is corresponding to the thermal magnetic irreversible temperature $T_{irr}$. The value of $R_S$ near the $T_N$ is about 0.3 cm$^{-3}$/C, which is comparable with that of some other antiferromagnets [15]. The temperature dependence of anomalous Hall conductivity (AHC) $\sigma_H$ is also shown in Fig. 2 (e). The comparation between $Cr_{0.68}Se$ and other materials with AHE is shown in **Table 3S (supporting materials)**. Figure 2 (f) presents the scaling behavior of $\rho_{xy}^A$ vs. $\rho_{xx}$. Note that there is a turning point at around 15 K, corresponding to the $T_{irr}$, as observed in the $\chi(T)$ curves. From the formula of $\rho_{xy}^A = \beta \rho_{xx}^\alpha$, the fits for 15 K $\leq T \leq$ 42 K and $T \leq$ 15 K gives the linear relation between $\rho_{xy}^A$ and $\rho_{xx}$, respectively. Both clearly indicate that the extrinsic skew-scattering mechanism dominates the AHE in the $Cr_{0.68}Se$ single crystal rather than the intrinsic *KL* mechanism or the extrinsic side-jump which gives the quadratic relationship between $\rho_{xy}^A$ and $\rho_{xx}$ [7, 8].

Let's try to understand the origin of AHE in $Cr_{0.68}Se$. Conventionally, there are three kinds of mechanisms dominate the AHE. The schematics are shown in Figs. 3 (a)-(c), for the *KL* mechanism, which shows that electrons obtain an anomalous group velocity from external electrical field, resulting to the AHE: [7]

$$\dot{\vec{r}} = \frac{\partial \vec{E}}{\hbar \partial \vec{k}} + \frac{e}{\hbar} \vec{E} \times \vec{b_n} \qquad (7)$$

where the second term of right side is the additional term induced by electrical field, it also can be seen as a Berry curvature term, which is an intrinsic property of the occupied electronic states in a given crystal with certain symmetry. It takes nonzero value only in systems where time-reversal symmetry is broken (for example, by applying a magnetic field) or where net magnetic moments present. It gives the scaling behavior $\rho_{xy}^A = \beta\rho_{xx}^2$. For the side-jump mechanism, which exhibits that the potential field induced by impurities contributes to the anomalous group velocity, it follows the same scaling behavior with the *KL* mechanism. However, for the skew-scattering mechanism, which shows that asymmetric scattering induced by impurity or defect contributes to the AHE, it presents the scaling behavior $\rho_{xy}^A = \beta\rho_{xx}$. All three kinds of mechanism are subjected to SOC. It is well established that the AHE displayed by ferromagnet with its magnetization [7]. Therefore, because of no net spin magnetization, the AHE is zero in conventional antiferromagnets with collinear magnetic moments, and from symmetry analysis, it also does not exist Berry curvature. However, some antiferromagnets with non-collinear AFM spin structure can also get rise to AHE by arising a Berry curvature [38, 39]. For our studied $Cr_{0.68}Se$, as shown in Fig. 3 (d), below $T_N$, the magnetic moments of $CrSe_2$ layer in $Cr_{0.68}Se$ tend to non-collinear AFM configuration, and the AHE occurs. To confirm the conclusion that the observation of AHE in $Cr_{0.68}Se$ system may originate from the non-collinear AFM spin structure, let's focus on the *M(H)* data, as shown in the inset of Figs. 1 (a) and (b). It is clear that $Cr_{0.68}Se$ has a nearly zero magnetic moment below the $T_N$, like $Mn_3Ge$ with the non-collinear AFM ground state and the feature of AHE below the $T_N$ [14]. Hence, we also can conclude that the AHE depicted in Fig. 2 (b) is related to the non-collinear AFM spin structure of $Cr_{0.68}Se$ [38-40]. On the other hand, as shown in Fig. 2 (f), the scaling behavior shows an anomaly at the $T_{irr}$. Figs. 3 (e) and (f) show the possible magnetic structure along *b* axis. There are two Cr atoms sites with different magnetic ordering. As the discussion above, the magnetic moments of intercalated Cr ions tend to be a PM arrangement above $T_{irr}$. Below the $T_{irr}$, they become to be canted ferromagnetism and be frozen and the magnetic coupling of the $CrSe_2$ layers is also enhanced along *c* axis. Thus, the irreversibility of ZFC and FCC curves result from the competition between the magnetic coupling of $CrSe_2$ layers and the intercalated Cr ions. In other word, the net magnetization of $Cr_{0.68}Se$ will be changed below $T_{irr}$. Therefore, the magnetism of intercalated Cr ions could have the influence to the AHE, leading to change the slope of the scaling behavior. In short, the CrSe$_2$ layer of Cr$_{0.68}$Se has the non-collinear AFM magnetic structure, while the intercalated Cr ions exhibits weak ferromagnetism with spin canting. They all have contribution to the observed AHE. Figure 2 (f) exhibits that the scaling behavior is still linear above and below $T_{irr}$ except for the slope of curve has been changed, which may indicate that the magnetic ordering of intercalated Cr ions is not the key role of the observed AHE and the origin of the AHE in Cr$_{0.68}$Se arises from the non-collinear AFM magnetic structure of Cr$_{0.68}$Se [38-40]. However, more experiments at low temperature, such as the neutron scattering experiments on powders and single crystals of Cr$_{0.68}$Se, are needed to determine the magnetic ground state in future.

In summary, we systematically investigate the magnetic, electronic, and thermal properties of the non-collinear AFM semiconductor Cr$_{0.68}$Se. A large negative MR is observed and may result from the non-collinear AFM magnetic structure of CrSe$_2$ layers in Cr$_{0.68}$Se. Then, we extract the temperature dependence of ordinary Hall coefficient $R_0$ and anomalous Hall coefficient $R_S$ from Hall measurement and find that Cr$_{0.68}$Se exhibit AHE under $T_N$ = 42 K, and the scaling behavior $\rho_{xy}^A = \beta \rho_{xx}$ reveals that it is the extrinsic skew-scattering mechanism rather than the intrinsic KL mechanism or the extrinsic side-jump mechanism that give rise to the AHE. And the AHE originates from the non-collinear AFM magnetic structure of CrSe$_2$ layers in Cr$_{0.68}$Se and the intercalated Cr ions influence the slope of the scaling behavior due to the strong magnetic coupling between CrSe$_2$ layers below $T_{irr}$. Our observation is helpful to find out other AFM semiconductors with AHE character, which may be useful in spintronic devices applications.

See supporting materials for the experimental details, X-ray diffraction and fitting parameter, energy dispersive spectroscopy (EDS), the Seebeck coefficient and comparison of some AHE parameters for some ferromagnets and antiferrmagnets.

## Acknowledgements

The work was supported by the National Key Research and Development Program under contract 2016YFA0401803 and the Joint Funds of the National Natural Science Foundation of China and the Chinese Academy of Sciences' Large-Scale Scientific Facility under contract U1432139, the National Nature Science Foundation of China under contracts 11674326, Key

**Figure 1:**

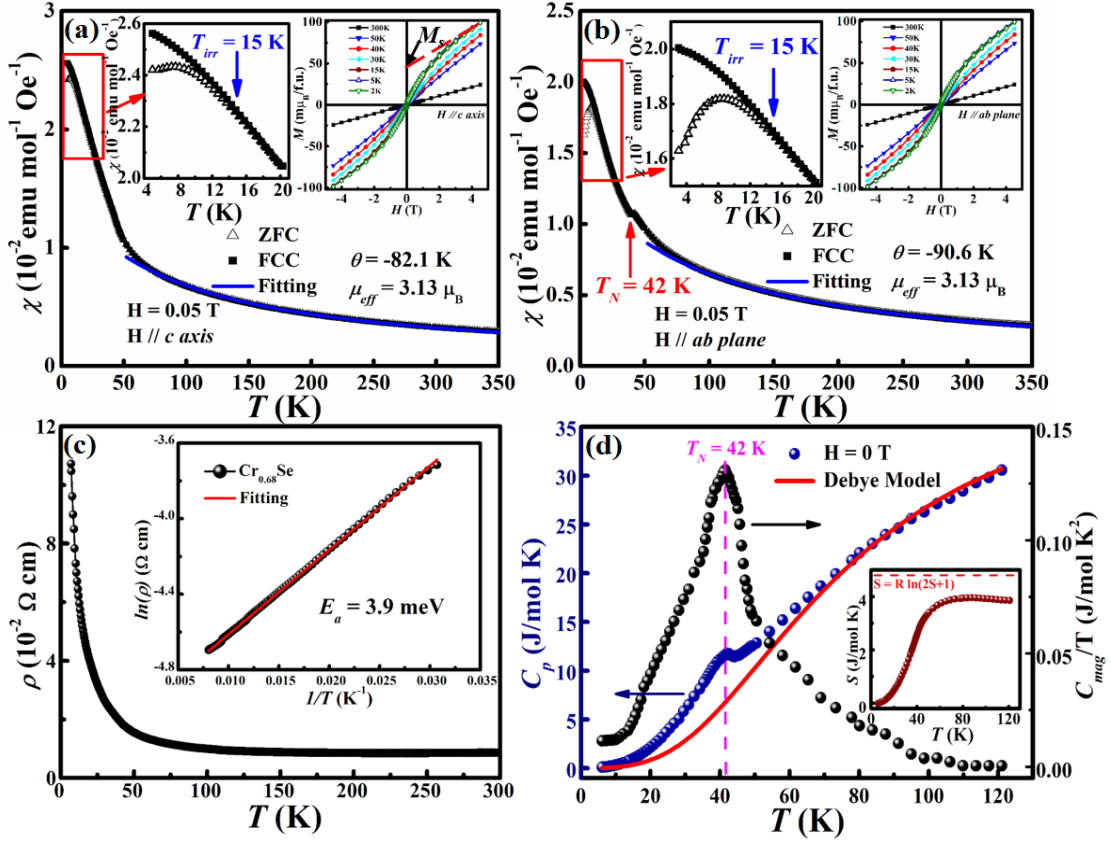

**Fig. 1 (color online):** The physical properties of $Cr_{0.68}Se$ single crystal. **(a)** and **(b)** The temperature dependence of magnetization of $Cr_{0.68}Se$ single crystal under *ZFC* and *FC* models with applied magnetic field *H*=0.05 T; **(a)** for applied magnetic field parallel to the *c* axis and **(b)** for the applied magnetic field parallel to *ab* plane. The blue solid line show the fitting results according to the Eq. (1). The insets of left side are the enlarged picture in low temperature. The insets of right side are the field dependence of magnetization *M(H)* for *H // c axis* and *H // ab plane*, respectively; **(c)** The resistivity as a function of temperature. The inset shows the fitting results according to the thermal active model; **(d)** temperature dependent $C_p$ and $C_{mag}/T$ under the *H* = 0 T. The solid line is the fitting line according the Debye model. In the inset of **(d)** shows the calculated magnetic entropy $S_{mag}$ *vs. T*. The red dash line is the theoretical expected value.

**Figure 2:**

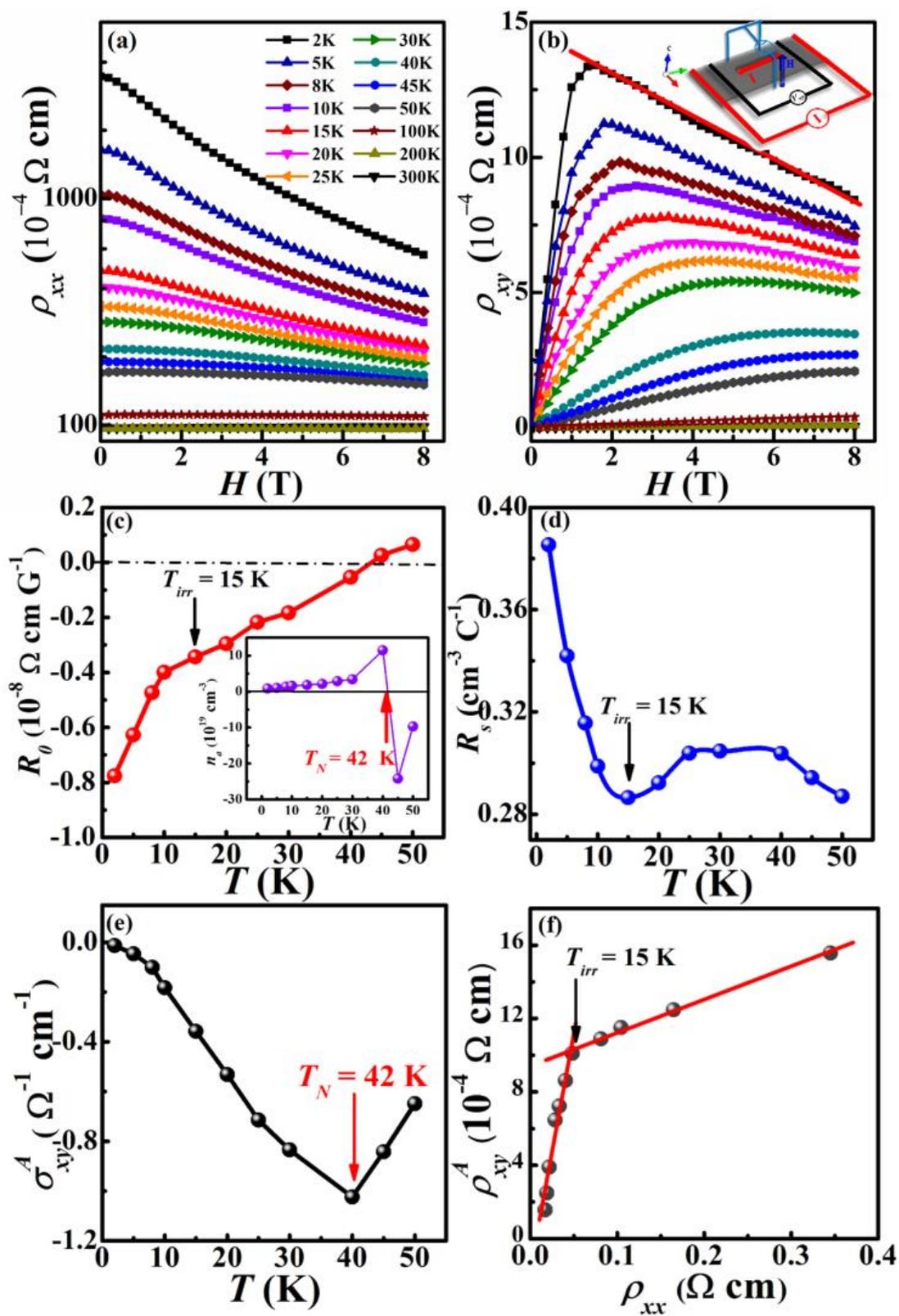

**Fig. 2 (color online):** The Hall measurements of $Cr_{0.68}Se$ single crystal. **(a)** The longitudinal resistivity $\rho_{xx}(H)$ as a function of $H$ at various temperatures; **(b)** Hall resistivity $\rho_{xy}(H)$ for the $Cr_{0.68}Se$ at various temperatures with $H//c$. The red solid line in **(b)** is the linear fit of $\rho_{xy}(H)$ at high-$H$ region when $T = 2$ K. Inset of **(b)** presents the schematic of Hall measurement; **(c)** and **(d)** The temperature dependence of fitted $R_0(T)$ and $R_s(T)$ from $\rho_{xy}(H)$ curves according to Eq. (6). Inset of **(c)** presents the derived $n_a(T)$ from $R_0(T)$; **(e)** Anomalous Hall conductivity $\sigma_{xy}^A(T)$ as a function of $T$; **(f)** The plot of $\rho_{xy}^A(T)$ vs. $\rho_{xx}(T)$. The red solid line are the fits by the relationship $\rho_{xy}^A \propto \beta\rho_{xx}^\alpha$ below and above 15 K, respectively.

**Figure 3:**

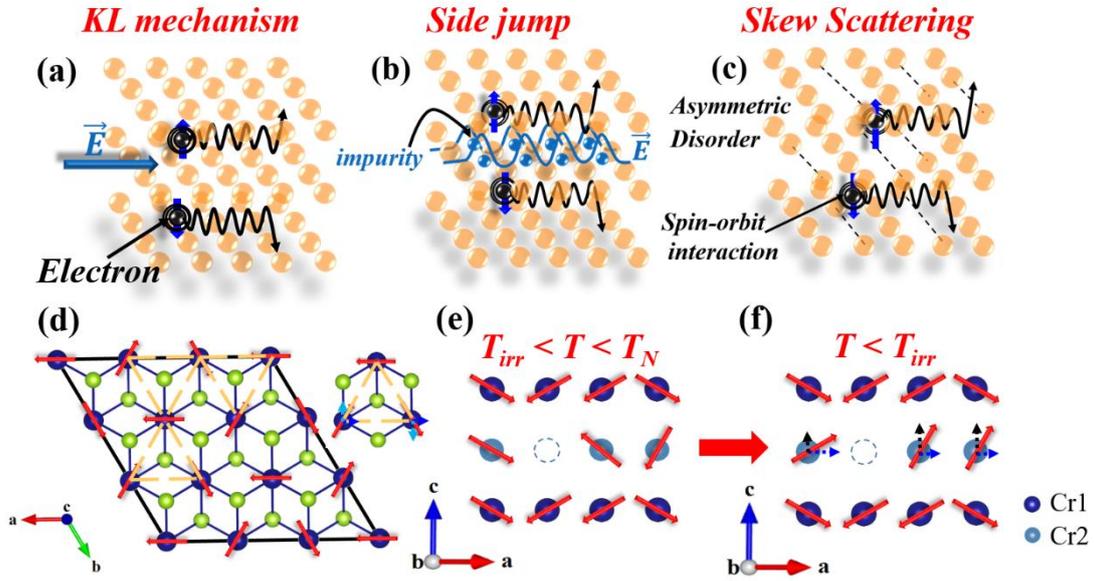

**Fig. 3 (color online):** Schematics for three mechanisms which are related to AHE. **(a)-(c)** The intrinsic *KL* mechanism, the extrinsic *side jump* mechanism, the extrinsic *skew-scattering mechanism*, respectively; **(d)** The possible non-collinear magnetic structure of the CrSe$_2$ layers in Cr$_{0.68}$Se along the *ab* plane; **(e)** and **(f)** The possible magnetic structure along *b* axis in Cr$_{0.68}$Se above $T_{irr}$ and below $T_{irr}$, respectively. Only the Cr atoms are shown, with their moments represented by arrows.